\documentclass[twocolumn]{bmcart}

\usepackage{amsthm,amsmath}
\RequirePackage{hyperref}
\usepackage[utf8]{inputenc} 
\usepackage{bm}
\usepackage{graphicx}
\usepackage{subfigure}
\usepackage{float}

\newcommand\scalemath[2]{\scalebox{#1}{\mbox{\ensuremath{\displaystyle #2}}}}

\startlocaldefs
\endlocaldefs

\begin{document}

\begin{frontmatter}

\begin{fmbox}
\dochead{Research}


\title{Clinical connectivity map for drug repurposing: using laboratory tests to bridge drugs and diseases}


\author[
   addressref={aff1},                   
   noteref={n1},                        
   email={wen.404@osu.edu}   
]{\inits{QW}\fnm{Qianlong} \snm{Wen}}
\author[
   addressref={aff2},
   noteref={n1},                        
   email={liu.7324@osu.edu}
]{\inits{RL}\fnm{Ruoqi} \snm{Liu}}
\author[
   addressref={aff2,aff3},
   corref={aff2,aff3},  
   email={zhang.10631@osu.edu}
]{\inits{PZ}\fnm{Ping} \snm{Zhang}}


\address[id=aff1]{
  \orgname{Department of Electrical and Computer Engineering, The Ohio State University}, 
  \street{2015 Neil Ave}               %
  \postcode{43210}                                
  \city{Columbus}, 
  \state{Ohio},
  \cny{USA}                                    
}
\address[id=aff2]{%
  \orgname{Department of Computer Science and Engineering, The Ohio State University},
  \street{2015 Neil Ave}               %
  \postcode{43210}                                
  \city{Columbus}, 
  \state{Ohio},
  \cny{USA}            
}
\address[id=aff3]{%
  \orgname{Department of Biomedical Informatics, The Ohio State University},
  \street{1800 Cannon Drive}               %
  \postcode{43210}                                
  \city{Columbus}, 
  \state{Ohio},
  \cny{USA}            
}


\begin{artnotes}
\note[id=n1]{These authors contributed equally to this work.} 
\end{artnotes}



\begin{abstractbox}

\begin{abstract} 
\parttitle{Background} 
Drug repurposing, the process of identifying additional therapeutic uses for existing drugs, has attracted increasing attention from both the pharmaceutical industry and the research community. Many existing computational drug repurposing methods rely on preclinical data (e.g., chemical structures, drug targets), resulting in translational problems for clinical trials.
\parttitle{Methods}
In this study, we propose a clinical connectivity map framework for drug repurposing by leveraging laboratory tests to analyze complementarity between drugs and diseases. We establish clinical drug effect vectors (i.e., drug-laboratory test associations) by applying a continuous self-controlled case series model on a longitudinal electronic health record data. We establish clinical disease sign vectors (i.e., disease-laboratory test associations) by applying a Wilcoxon rank sum test on a large-scale national survey data. Finally, we compute a repurposing possibility score for each drug-disease pair by applying a dot product-based scoring function on clinical disease sign vectors and clinical drug effect vectors.
\parttitle{Results}
We comprehensively evaluate 392 drugs for 6 important chronic diseases (include asthma, coronary heart disease, congestive heart failure, heart attack, type 2 diabetes, and stroke). We discover not only known associations between diseases and drugs, but also many hidden drug-disease associations. For example, clopidogrel and alendronate may be repurposed as candidate drugs for diabetes and cardiovascular diseases respectively. Moreover, we are able to explain the predicted drug-disease associations via the corresponding complementarity between laboratory tests of drug effect vectors and disease sign vectors.
\parttitle{Conclusion}
The proposed clinical connectivity map framework uses laboratory tests from electronic clinical information to bridge drugs and diseases, which is explainable and has better translational power than existing computational methods. Experimental results demonstrate the effectiveness of the proposed framework and suggest that our method could help identify drug repurposing opportunities, which will benefit patients by offering more effective and safer treatments.
\parttitle{Availability} 
The code for this paper is available at: \url{https://github.com/HoytWen/CCMDR}
\end{abstract}




\end{abstractbox}
\end{fmbox}

\end{frontmatter}

\section*{Introduction}


Traditional \textit{de novo} drug discovery is a long and complicated process \cite{o2005finding, chong2007new}, which usually takes more than 15 years \cite{dimasi2001new}, and costs 800 million to 1 billion US dollars \cite{adams2006estimating} to develop a new drug. Drug repurposing, investigation of potential additional uses for existing drugs, is becoming an appealing research field given its potential in lowering overall costs and shortening drug development timelines \cite{pushpakom2019drug}.

There has been a surge of computational methods proposed for drug repurposing in recent years, which can be roughly classified into two categories based on different data sources: preclinical data-based and clinical data-based. Preclinical data-based methods often build machine learning models based on preclinical data, such as drug chemical structure, protein targets and gene expression information, to identify potential drug-disease associations. For example, Keiser et al. \cite{keiser2009predicting} use drug structural similarity as the measurements to find the drugs with similar effects. Lamb et al. \cite{lamb2006connectivity,lamb2007connectivity} raise the connectivity map (CMap) approach for drug repurposing by using gene expression data, which is based on molecular activity. Luo et al. \cite{luo2016dpdr} develop a server named DPDR-CPI which predicts the new indications of existing drugs by analyzing the chemical-protein interactome (CPI) profile. Some researchers also tried to construct computational frameworks that integrated several kinds of data sources and even disease similarity measurement profiles to make better predictions. PreDR model proposed by Wang et al. \cite{wang2013drug} integrated drug structure, drug target, side-effects and disease phenotype data to find the novel drug indications. Zhang et al. \cite{zhang2018predicting} raised a similarity constrained matrix factorization method to predict drug-disease association based on known drug-disease associations, drug features and disease semantic information. However, all of these methods rely heavily on preclinical information to make predictions. This will cause a large translation gap when we apply the drugs on humans. It is estimated that of all compounds effective in cell assays, only 30\% of them could work in animals and only 5\% of them could work in humans \cite{pammolli2011productivity}.

Compared with preclinical data, clinical data provide more applicable and reliable data sources for drug repurposing as clinical information (e.g., laboratory test results) may be seen as valuable read-outs of drug effects directly on human bodies. It is directly observed form patients, so there is no need to consider about the translational problems. Many computational frameworks based on clinical information has been raised due to the large amount of available electronica clinical data. 
Jung et al. \cite{jung2013inferring} find the connection between drugs and diseases in clinical diagnose notes by literature mining, but it does not include any other structured data, like laboratory test results. Jang et al. \cite{jang2016inferring} propose a framework that use laboratory test results to reflect the influence of drugs and diseases on human physiological activities, and the method they use to establish drug effects is counting co-occurrence between drug and laboratory tests. However, it is not efficient enough to dig the hidden relation between drugs and laboratory tests, especially when we have a large dataset and include many laboratory and existing drugs in our experiment. Kuang et al. \cite{kuang2016computational} and Ghalwash et al. \cite{ghalwash2017exploiting} raised more advanced methods to compute the influence of drugs on laboratory tests, however, they reflect the effect of drugs on single laboratory (e.g., blood sugar level), which it is not enough to represent the state of the complex human system. It would be more efficient and accurate if we build an electronic clinical information-based drug repurposing framework and implement it by more efficient statistical analysis methods designed for large datasets. During this process, we will include as many laboratory tests as we can in our experiment to completely represent the state of human biological system. The idea of CMap raised by Lamb et al. \cite{lamb2006connectivity,lamb2007connectivity} which uses gene expression values to bridge drugs and diseases, directly inspires us to formulate and leverage all the laboratory tests involved in our experiment to build associations between drugs and diseases from clinical perspective.

In this paper, we propose a clinical connectivity map framework for drug repurposing (CCMDR) by leveraging laboratory tests to analyze the influence of drugs and diseases on the human biological system. Specifically, we first establish clinical disease sign vectors (i.e., disease-laboratory test associations) by applying a Wilcoxon rank sum test on a large-scale national survey data. We then establish clinical drug effect vectors (i.e., drug-laboratory test associations) by applying a continuous self-controlled case series model on a longitudinal electronic health record data. Finally, we compute a repurposing possibility score for each drug-disease pair by applying a dot product-based scoring function on clinical disease sign vectors and clinical drug effect vectors. Experimental results show that our method can not only retrieve the known drug-disease associations in high accuracy but also can find potential indications, which can be verified from medical literature. For example, clopidogrel and alendronate may be repurposed as candidate drugs for diabetes and cardiovascular diseases respectively. Moreover, we can explain the predicted drug-disease associations via the corresponding complementarity between laboratory tests of drug effect vectors and disease sign vectors. So, it is suggested that our method can be potentially used in drug repurposing tasks.


In brief, the contribution of the paper can be summarized as below:
\begin{itemize}
\item We propose a clinical connectivity mapping framework for drug repurposing. The new framework solely based on the clinical patient data, thus with less translational problems.
\item We evaluate our framework for 392 drugs on 6 important chronic diseases (include asthma, coronary heart disease, congestive heart failure, heart attack, type 2 diabetes, and stroke). Experimental results show that our method achieves high accuracy in retrieving the known indications of drugs.
\item We study the predicted drug repurposing candidates via the corresponding complementarity between laboratory tests of drug effect vectors and disease sign vectors. Case studies with literature support show the potential of our method to discover previously unknown indications of existing drugs.
\end{itemize}

\begin{figure*}[h!]
\centering
\includegraphics[width=0.95\textwidth]{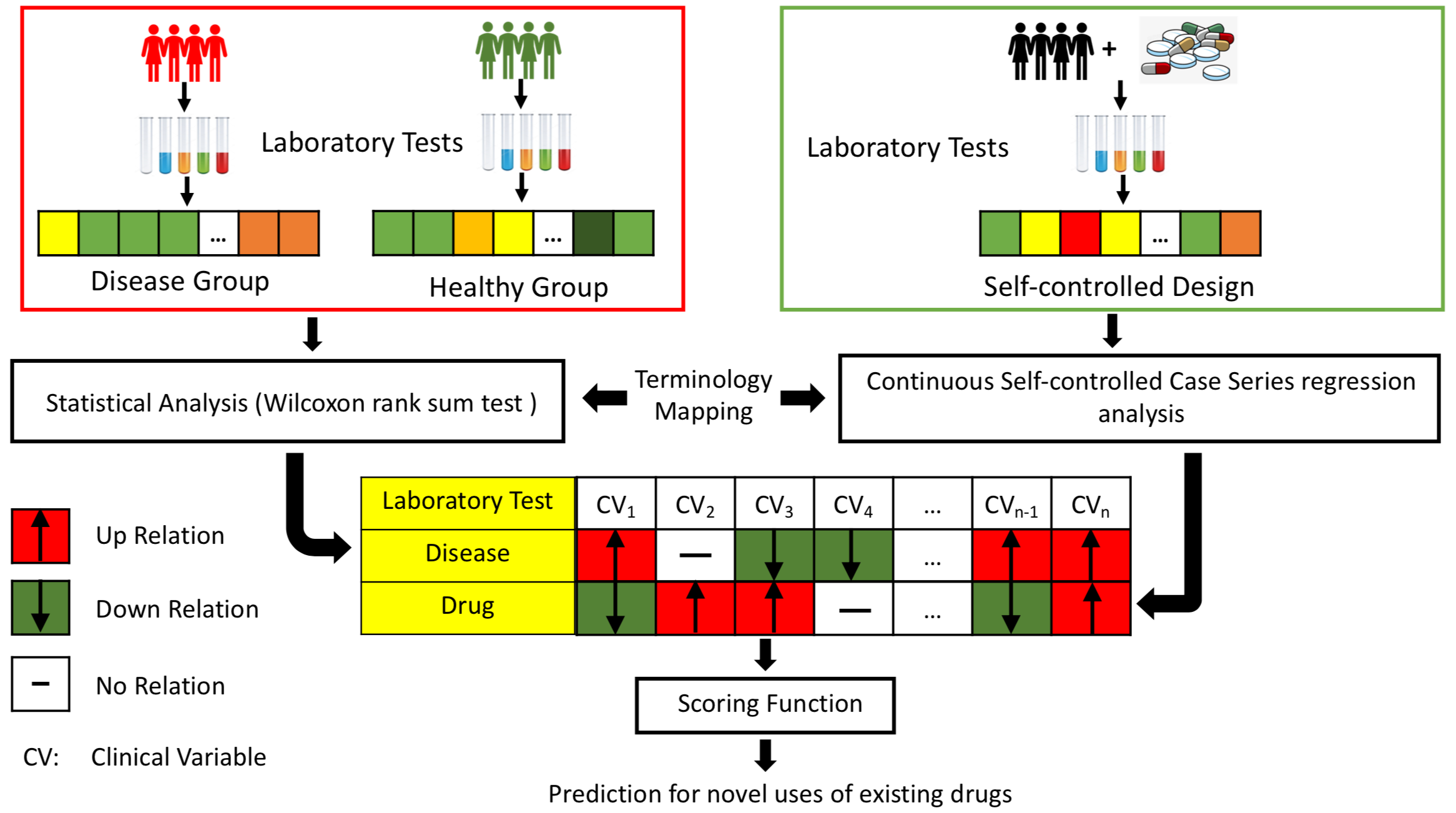}
\caption{This figure presents the pipeline of our framework. The framework contains three main components: (1) establishing clinical drug effect vectors by applying a continuous self-controlled case series model on a longitudinal electronic health record data (EHR), (2) establishing clinical disease sign vectors by applying a Wilcoxon rank sum test on a large-scale national survey data (NHANES), (3) computing repurposing possibility score for each drug-disease pair by applying a dot product-based scoring function on clinical disease sign vectors and clinical drug effect vectors. We do a terminology mapping before we establish the clinical drug effect vectors and clinical disease sign vectors to make sure each clinical vector includes the same laboratory tests. There are three kinds of relation types in the clinical vectors ("Up", "Down", "No"), which represent increasing, decreasing and not significantly changing laboratory tests level, respectively. }
\label{fig:idea}
\end{figure*}

\section*{Methodology}
\subsection*{Dataset and Data Preprocess }


We use the questionnaire and laboratory test results from the National Health and Nutrition Examination Survey (NHANES) \cite{cdc2005national} to establish the clinical disease sign vectors. According to the questionnaire survey (e.g., "Has been diagnosed with type 2 diabetes?"), individual samples are divided into disease group (who answered "yes") and healthy group (who answered "no"). Next, we perform the statistical analysis to identify those disease-related clinical variables from collected laboratory test results in NHANES data. We extract 87,464 individual samples, 986 numerical clinical variables and more than 30 disease conditions from NHANES data range from 1999 to 2016. Here, we only consider the disease conditions with more than 1000 individual samples, which results in 6 unique diseases (i.e., asthma, coronary heart disease, congestive heart failure, heart attack, type 2 diabetes and stroke).

We use the prescription and laboratory test result histories of patients in Electronic Health Record (EHR) to establish the clinical drug effect vectors. We transform the prescription records of patients into matrixes based on medication use situations. To study the associations between prescribed drugs and laboratory test results, we apply a continuous self-controlled case series model \cite{kuang2016computational} to analyze the effects of a drug on the laboratory test results. Our proprietary EHRs contain comprehensive health records of more than 300 thousand patients over four years. We only consider patients with complete records (i.e., having both prescription and its corresponding laboratory test results), which results in 91,934 patients, 1,344 kinds of treatments and 65 kinds of laboratory tests. After excluding those prescriptions with less than 1000 patients, we obtain 392 unique prescribed drugs.

We bridge the drug and disease using the laboratory test results obtained from each side. Since the laboratory test results are from different data resources (i.e., national survey data and electronic health records), we need to standardize those laboratory tests for further analysis. The laboratory tests that appear in both datasets are included and mapped to a standard list with consistent names. Also, the non-numerical laboratory tests are excluded. Finally, we obtain 35 laboratory tests considered as clinical variables. The full list of the 35 clinical variables can be found in \textbf{Table S1}. Our inference of a drug-disease pair is based on the complementary and adverse effects that each drug candidate and disease condition has on the 35 clinical variables.

\subsection*{Clinical Disease Sign Vector}
We extract 6 disease conditions and 35 clinical variables from NHANES after preprocessing to establish the clinical disease sign vector.
The dimension of each disease sign vector is $1 \times 35$.  There are three types of relations between a disease and clinical vectors (i.e., ”Up”, ”Down” and ”No”), which represents increasing, decreasing and not significantly changing of laboratory tests level, respectively. As mentioned above, the combined data is divided into disease group and control group according to the questionnaire data, we apply Wilcoxon rank sum test (a.k.a., Mann Whitney U test) on two groups to calculate the p-value for each clinical variable. To get the p-value, we combine the values of the two groups and rank them. Then we calculate two statistical values $U_{1}$ and $U_{2}$, which are defined as follows:
\begin{equation}
\begin{array}{c}{U_{1}=R_{1}-\frac{n_{1}(n_{1}+1)}{2}} \\ {U_{2}=R_{2}-\frac{n_{2}(n_{2}+1)}{2}}\end{array}\label{1}
\end{equation}
where $n_{1}$ and $n_{2}$ are the sample sizes of the two groups, $R_{1}$ and $R_{2}$ are the sum of the ranks in the two groups, respectively. The smaller value of $U_{1}$ and $U_{2}$ is used to consult the Mann-Whitney significance table. Certain p-value cut-off is used to examine whether the value change is significant or not \cite{storey2003statistical}. In our work, the p-value threshold is set to be 0.05. Only the clinical variables satisfy the condition that p-values are less than 0.05 can be regarded as significant clinical variables concerning the disease. We consult the Mann-Whitney table of $\alpha=0.05$. If the smaller value of $U_{1}$ and $U_{2}$ is larger than the value given in the table, the null hypothesis is true otherwise false. Then we assign relation direction to this clinical variable by comparing the average clinical variable value of the disease group and control group. Up relation ("$\uparrow$") indicates a significant value increase in the disease group compared with the control group, while down relation ("$\downarrow$") means the laboratory test value of the disease group is significantly lower than that of the control group, no relation ("-") indicate the laboratory test level will not be significantly influenced by the disease.

\subsection*{Clinical Drug Effect Vector}

To establish clinical drug effect vectors, we extract 392 drugs and 35 clinical variables from EHR data. The clinical variables used here are the same as ones in establishing the disease sign vectors. So, the dimension of each drug effect vector is $1 \times 35$. We need to consider the prescription records of patients and their corresponding laboratory test results records simultaneously, and the EHR dataset we use is a large dataset that includes millions of records. So, we need to find a way to analyze the high-dimensional longitudinal data. In our work, we adopt the continuous self-controlled case Series (CSCCS) model proposed by Kuang et al. \cite{kuang2016computational}, it is a lasso regression analysis model designed to do the data analytical work for EHR dataset. 

Assuming there are N patients with a specific kind of clinical variable measurement and M kinds of drugs in EHR dataset. Continuous variable $y_{i j}$, where $i \in\{1,2, \cdots, N\}$, $j \in\{1,2, \cdots, J_{i}\}$, indicates the value of $j_{t h}$ clinical variable measurement taken among a total number of $J_{i}$ measurements for the $i_{t h}$ patient, while binary variable $x_{i j m}$, where $i \in\{1,2, \cdots, N\}$, $j \in\{1,2, \cdots, J_{i}\}$, $m \in\{1,2, \cdots, M\}$, are used to indicated the drug whether $i_{t h }$ patient are exposed to the $m_{t h}$ drug when the $j_{t h }$ clinical variable measurement is taken. 0 represents no and 1 represents yes. 

$y_{i j}$ is regard as the output variables when we fit the structured data into the linear regression model, so we have:

\begin{equation}
y_{i j} | \boldsymbol{x}_{i j}=\alpha_{i}+\boldsymbol{\beta}^{\top} \boldsymbol{x}_{i j}+\epsilon_{i j}, \quad \epsilon_{i j} \stackrel{i i d}{\sim} N\left(0, \sigma^{2}\right)\label{2}
\end{equation}
\centerline{$
\boldsymbol{\beta}=
\begin{bmatrix}
\beta_{1} & \beta_{2} & \cdots & \beta_{M}
\end{bmatrix}^{\top},$}
\centerline{$
\boldsymbol{x}_{i j}=
\begin{bmatrix}
x_{i j 1} & x_{i j 2} & \cdots & x_{i j M}
\end{bmatrix}^{\top}
$}

$\alpha_{i}$  in equation (\ref{2}) represents the average baseline level of $y_{i j}$ on $i_{t h}$ patient. That means it is independent of the date the measurement was taken and drugs the patient used when the measurement was taken. Each patient has an individual baseline value. $\epsilon_{i j}$ here is an independent and identically distributed Gaussian noises with zero means and fixed but unknown variance $\sigma^{2}$. Then the linear model can be easily converted to a least square problem as follows:

\begin{equation}
\scalemath{0.9}{
\arg\min_{\boldsymbol{\alpha},\boldsymbol{\beta}}\mathcal{L}(\boldsymbol{\alpha},\boldsymbol{\beta})=\arg\min_{\boldsymbol{\alpha},\boldsymbol{\beta}}\frac{1}{2}
\begin{Vmatrix}
\boldsymbol{y}-
\begin{bmatrix}
\boldsymbol{Z} & \boldsymbol{X}
\end{bmatrix}
\begin{bmatrix}
\boldsymbol{\alpha}\\
\boldsymbol{\beta}
\end{bmatrix}
\end{Vmatrix}^{2}_{2}
\label{3}
}
\end{equation}
where\\ 
$\boldsymbol{\alpha}=
\begin{bmatrix}
\alpha_{1} & \alpha_{2} & \cdots & \alpha_{N}
\end{bmatrix}^{\top}$, $\boldsymbol{Z}=\operatorname{diag}
\begin{pmatrix}
\mathbf{1}_{1}, \cdots, \mathbf{1}_{N}
\end{pmatrix}$,\\
$\boldsymbol{y}=
\begin{bmatrix}
y_{11} & \cdots & y_{1 J_{1}} & {\cdots} & y_{N 1} & \cdots & y_{N J_{N}}
\end{bmatrix}^{\top}$,\\
$\boldsymbol{X}=
\begin{bmatrix}
\boldsymbol{x}_{11} & \cdots & \boldsymbol{x}_{1 J_{1}} & \cdots & \boldsymbol{x}_{N 1} & \cdots & \boldsymbol{x}_{N J_{N}}
\end{bmatrix}^{\top}$ \\
where $\boldsymbol{Z}$ is a block diagonal matrix and $\boldsymbol{1_{i}}$ is a $J_{i} \times 1$ vector in which all the components are 1. By solving this problem, we can get the optimized parameter $\boldsymbol{\beta}$, which is also the interest of our task. $\boldsymbol{\beta}$ is a $1 \times M $ parameter vector, parameter $\boldsymbol{\beta_{m}}$ in $\boldsymbol{\beta}$ indicates the effect of $m_{t h}$ drug on the output variable $\boldsymbol{y}$. The optimized parameter we get with the CSCCS model is numerical. Positive and negative parameters in this vector represent the corresponding drugs that may increase and decrease the level of output variable respectively, while 0 indicates the corresponding drugs do not influence it. In the CSCCS model, parameter $\boldsymbol{\alpha}$ is regarded as a nuisance parameter, our interest is parameter $\boldsymbol{\beta}$ so we do not need to care the value of $\boldsymbol{\alpha}$. To eliminate the effect of $\boldsymbol{\alpha}$, \cite{kuang2016computational} consider:

\begin{small}
\begin{equation}
\scalemath{0.8}{
\frac{\partial \mathcal{L}(\boldsymbol{\alpha}, \boldsymbol{\beta})}{\partial \boldsymbol{\alpha}}=\mathbf{0} \Rightarrow \boldsymbol{\alpha}=\left(\boldsymbol{Z}^{\top} \boldsymbol{Z}\right)^{-1} \boldsymbol{Z}^{\top}(\boldsymbol{y}-\boldsymbol{X} \boldsymbol{\beta})=\overline{\boldsymbol{y}}-\overline{\boldsymbol{X}} \boldsymbol{\beta}
}
\label{4}
\end{equation}
\end{small}

Where $\overline{\boldsymbol{y}}$ is a $N\times 1$ vector which includes the average value of clinical value among $N$ patients, $\overline{y_{i}} = \frac{1}{J_{i}} \sum_{j=1}^{J_{i}} y_{i j}$. $\overline{\boldsymbol{X}}$ is a $N \times M$ matrix and $\overline{\boldsymbol{X}}_{i}=\frac{1}{J_{i}} \sum_{j=1}^{J_{i}} \boldsymbol{x}_{i j}^{\top}$. So, the expression of CSCCS model below, which is free of $\boldsymbol{\alpha}$, is derived by substituting equation (\ref{4}) into equation (\ref{3}):

\begin{equation}
\arg \min _{\beta} \frac{1}{2}\|\boldsymbol{y}-\boldsymbol{Z} \overline{\boldsymbol{y}}-(\boldsymbol{X}-\boldsymbol{Z} \overline{\boldsymbol{X}}) \boldsymbol{\beta}\|_{2}^{2}\label{5}
\end{equation}

When we apply the CSCCS model on the high-dimensional longitudinal EHR data, we will add a $L_{1}$ penalty term because there is an assumption that the level of clinical variables will only be significantly influenced by a small portion of drugs. The $L_{1}$ penalization drives most components of $\boldsymbol{\beta}$ to zero or closed to zero \cite{tibshirani1996regression}. In other words, we simply want to know the drugs which are most correlated to the level change of clinical variables. So the final expression of the CSCCS model we apply to this problem is:

\begin{equation}
\arg \min _{\boldsymbol{\beta}} \frac{1}{2}\|\boldsymbol{y}-\boldsymbol{Z} \overline{\boldsymbol{y}}-(\boldsymbol{X}-\boldsymbol{Z} \overline{\boldsymbol{X}}) \boldsymbol{\beta}\|_{2}^{2}+\lambda\|\boldsymbol{\beta}\|_{1}\label{6}
\end{equation}
where $\lambda>0$, $\lambda$ decides the sparsity of optimized result so we need to tune this parameter to get a final result with proper sparsity level.

In order to further filter out the drugs which do not have a significant effect on clinical variables, our implementation also returns the p-value of each component in $\boldsymbol{\beta}$. We apply the same p-value cut-off strategy on the optimized result. Parameters with a p-value greater than 0.05 in $\boldsymbol{\beta}$ are regarded as insignificant effect and we assume their corresponding drugs are uncorrelated with clinical variable level change. The significant effects can be divided into increasing or decreasing effect based on the coefficient value is positive or negative. Then we assign each drug-clinical variable pair up ("$\uparrow$"), down ("$\downarrow$") and no ("-") relation type just like clinical disease sign vectors.


\subsection*{Scoring Function}

After we establish the clinical vectors for each drug-disease pair, we need to define a scoring function to calculate the repurposing possibility score for each drug-disease pair. The inference for each drug-disease pair is based on complementary and adverse effects. Specifically, complementary effect refers to the opposed relation type between a clinical disease sign vector and clinical drug effect vector on the same clinical variables, while adverse effect refers to the same relation type between a clinical disease sign vector and clinical drug effect vector on the same clinical variables. The complementary relation direction between the two vectors will increase the final repurposing possibility score of a drug-disease pair while adverse relation direction will decrease it. Here, we use a dot product-based scoring function to consider both complementary and adverse effects of a drug candidate on a disease. The scoring function can be written as follow: 
\begin{equation}
TS_{disease-drug}=-CV_{Drug} \cdot CV_{Disease}\label{7}
\end{equation}
where $CV_{Drug}$ is the clinical drug effect vector, and $CV_{Disease}$ is the clinical disease sign vector. We transform the 3 kinds of relation type in clinical vectors ("$\uparrow$", "$\downarrow$" and "-") into numerical values (1.0, -1.0, 0.0) for the convenience of calculation. To rank the drugs in descending order and emphasize the most powerful drug candidates predicted by our model, we add a minus sign before the product of the two vectors. So, the positive result calculated by this scoring function means there are more complementary relation directions than adverse relation directions between a drug candidate and a disease, while negative results indicate more adverse relation directions between this pair.

\section*{Results and Discussion}
\subsection*{Evaluation Metrics}
After we calculate the repurposing possibility score of each drug-disease pair, we need to prove that the score is qualified enough to serve as a metric to show whether a drug candidate is likely to be the potential treatment or not. The final drug candidate list is sorted by the repurposing possibility score in a descending order for the convenience of validation. The validation data we use comes from Side Effect Resource(SIDER) \cite{kuhn2015sider}. It contains drugs with indications or side-effects for many kinds of disease conditions. We take it for ground truth to testify whether our method can retrieve known indications of drugs. The hypothesis is that drug candidates with higher repurposing possibility score are more likely to be the treatment of the disease, which means most of the top-ranking drugs can be found in the drugs with indication and most of the bottom ranking drugs can be found in the drugs with side-effects provided by SIDER. In this case, those drugs can not be found in the validation data but still predicted with high repurposing possibility score by our model could be served as a potential treatment of the disease. So we need to use some evaluation metrics to test whether the known drug-diseases pairs are enriched at the top of our prediction list. We will use two kinds of evaluation metrics to validate our prediction. 

\subsubsection*{Precision at K}
The First kind of evaluation metric is precision at K. The top K precision value is the ratio of known treatment for a disease among the top K drug candidates for the disease predicted by our framework.
For each disease, we rank the drugs using the calculated repurposing possibility score. Then we compute the precision at $K$ ($K \in \{5, 10, 15, 20\}$) of each disease using the top-ranked $K$ drugs (e.g., precision at 10 corresponds to the proportion of correct retrieved drugs among the top 10 ranked drugs).


\subsubsection*{Fold-Enrichment Test}
Another evaluation metric to access whether our repurposing possibility score is correlated with the likelihood that disease-drug pair occurs or not is the fold-enrichment (FE) test. FE score can be defined by the following formula:
\begin{equation}
\text {FE Score}=\frac{(n / m)}{(N / M)}\label{8}
\end{equation}
where $M$ is the number of all the mapped drugs and $N$ is the number of drugs in the gold-standard dataset corresponding to each kind of disease condition. We will divide all the mapped drugs evenly into several groups according to their repurposing possibility score. So, $m$ is the total number of drugs in one group and $n$ is the number of drugs involved in the gold-standard dataset within the group. FE test can demonstrate the enrichment of known disease-drug pair (we assume the drug-disease pairs in SIDER is ground truth) within different score ranges. Our prediction can be proved to be reasonable if the FE test score is positively correlated with the repurposing possibility score. There are 392 drug-disease pairs for each kind of disease condition in our experiment, and all of them are ranked by repurposing possibility score and binned into groups of 80 pairs (the last group contains 72 drug-disease pairs). The scoring function is reasonable if the FE score is decreasing with the ascending order of the 5 groups because the average repurposing possibility score of each group is decreasing in that order.


\subsection*{Established Disease and Drug Vectors}
In our experiment, we first establish all of the clinical diseases and drug vectors. All the clinical disease sign vectors are represented in \textbf{Table S2}, and all the clinical drug effect vectors are represented in \textbf{Table S3}. 



\begin{figure*}[h!]
\centering
\centerline{\includegraphics[scale=0.45]{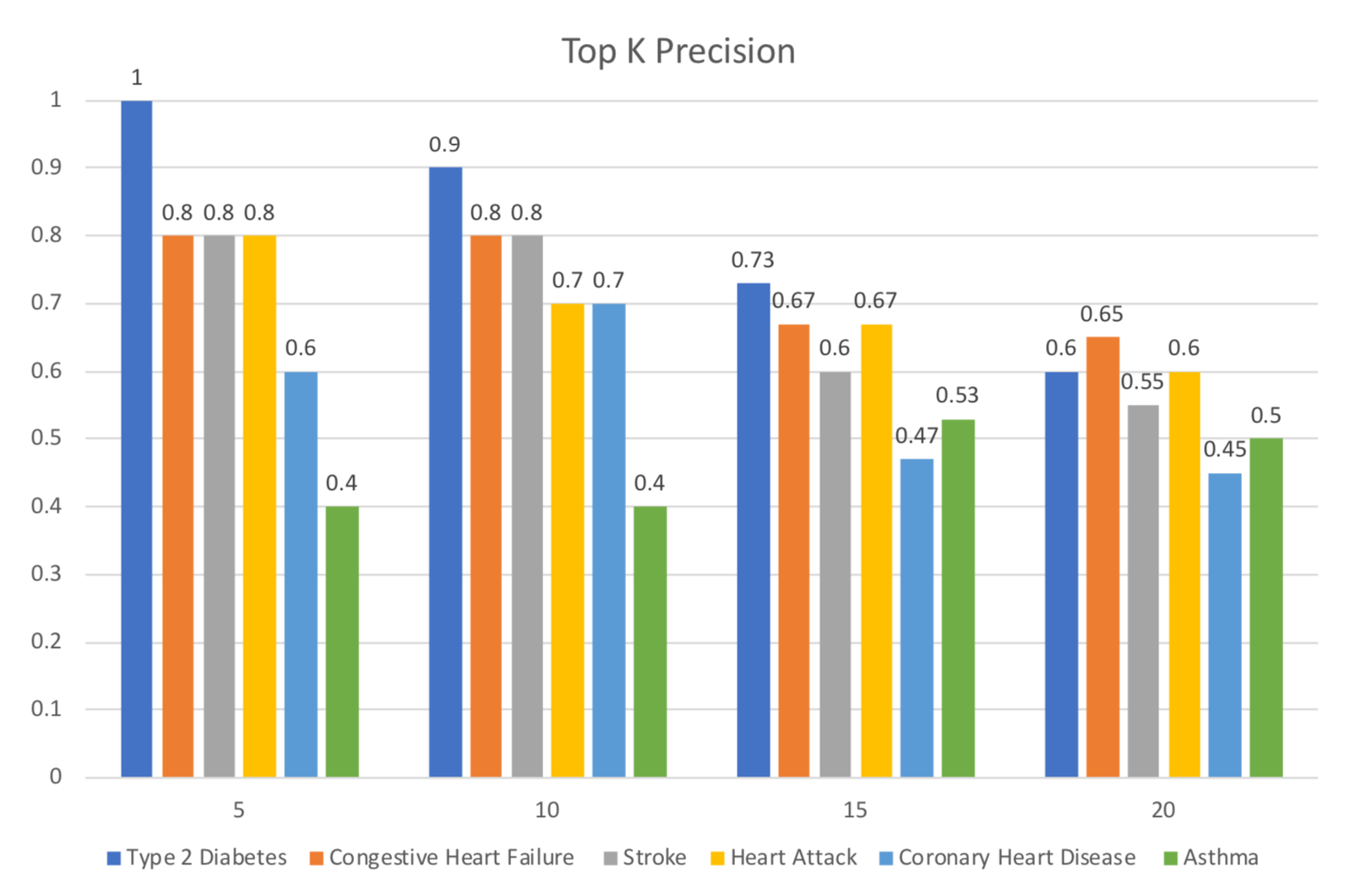}}
\caption{
Top K precision of each disease condition demonstrates the proportion of known drug-disease pairs among the top K ranking drugs in our prediction list. This prediction list is ranked by repurposing possibility score.
}
\label{fig:top-k}
\end{figure*}

\begin{figure*}[h!]
\centering
\centerline{\includegraphics[scale=0.7]{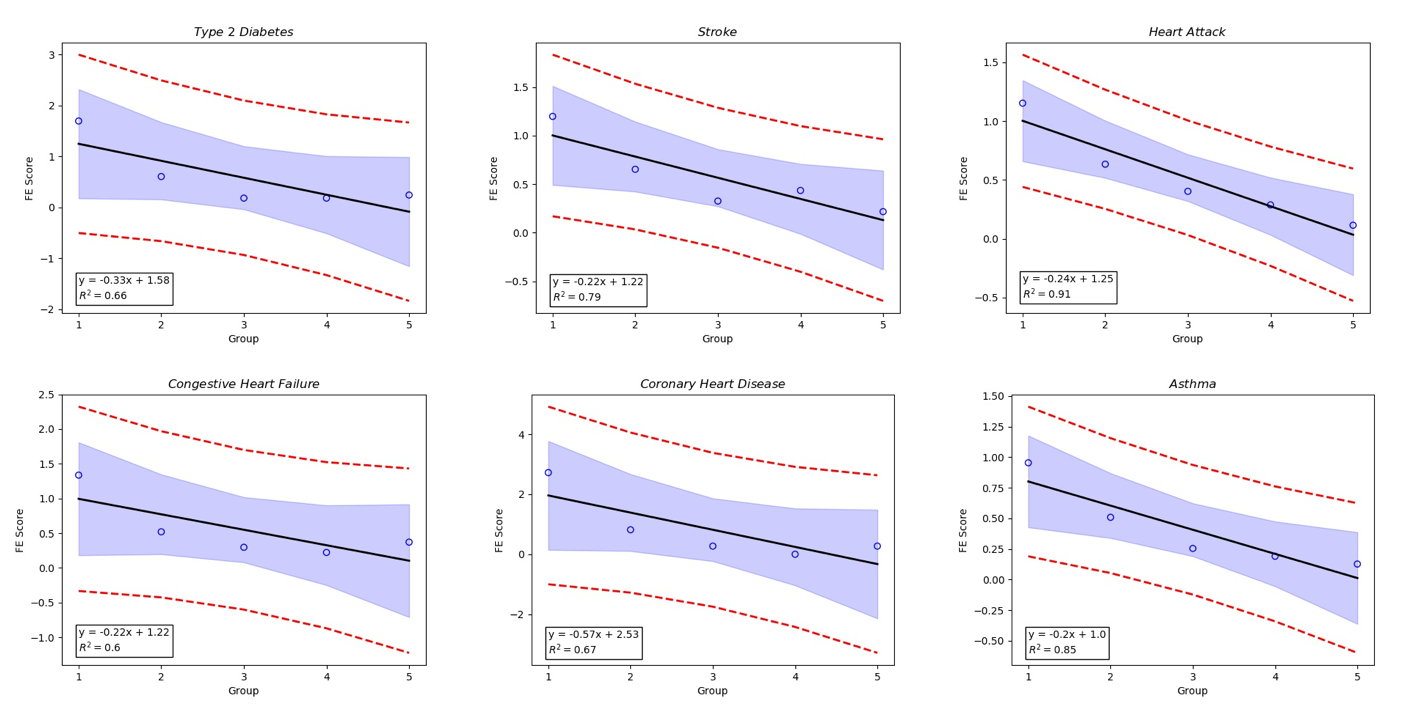}}
\caption{Fold-enrichment result of each disease condition, we divide the drug-disease pair into 5 groups with the descending order of average repurposing possibility score. So, the negative linear relations between the group order and FE score indicate the positive relationship between the average repurposing possibility score. It shows our scoring function is useful in finding the drugs which have a therapeutic effect on target diseases.}
\label{fig:FEScore}
\end{figure*}

\begin{figure}[!h]
\centering
\centerline{\includegraphics[width=0.95\linewidth]{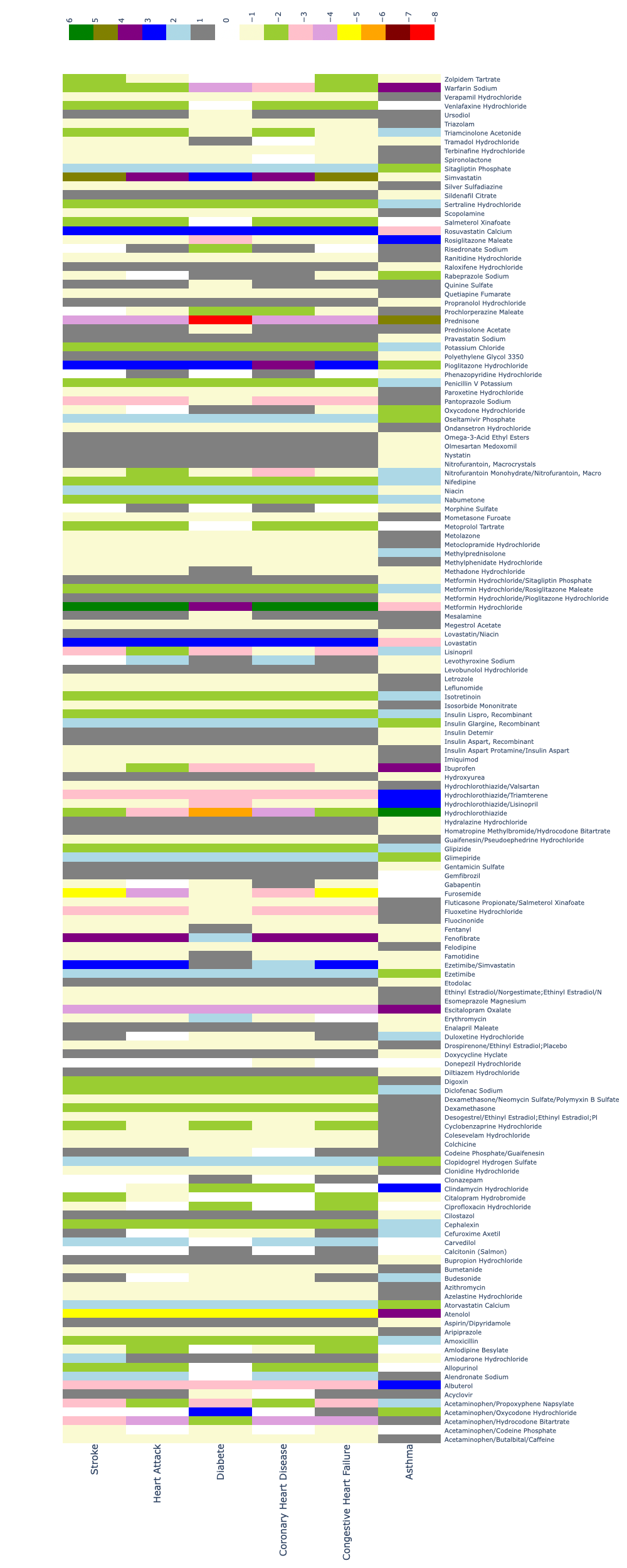}}
\caption{Heat map of drug-disease repurposing possibility scores. X-axis stands for the 6 disease conditions and Y-axis is the name of the 153 drugs that have an influence on any of the diseases involved in our experiment. The color bar above the heat map annotates the scores that different colors in the heat map stand for.}
\label{fig:DDS}
\end{figure}

Then, we calculate the repurposing possibility score of 392 kinds of drugs on six disease conditions (asthma, coronary heart disease, congestive heart failure, heart attack, type 2 diabetes and stroke). The repurposing possibility score of each drug-disease pair is listed in \textbf{Table S4}. We also transform the table into a heat map Figure \ref{fig:DDS} to vividly present the repurposing possibility score. Due to page limitation, we just present the drugs that have an influence on any of the 6 diseases in our experiment (153 kinds of drugs). The complete heat map can be is in \textbf{Figure S3}. Then we perform validations on our prediction for each of the six disease conditions. Each of the six disease conditions has enough sample size which can make our validation result more confident. We extract a list of drugs from the drug indication information resources for each of the six disease conditions provided by SIDER. All of the drugs in the six lists are known to treat the six disease conditions respectively, so we assume them as the ground truth and further compare them with our prediction. 

\begin{figure}[!h]
\centering
\centerline{\includegraphics[width=0.95\linewidth]{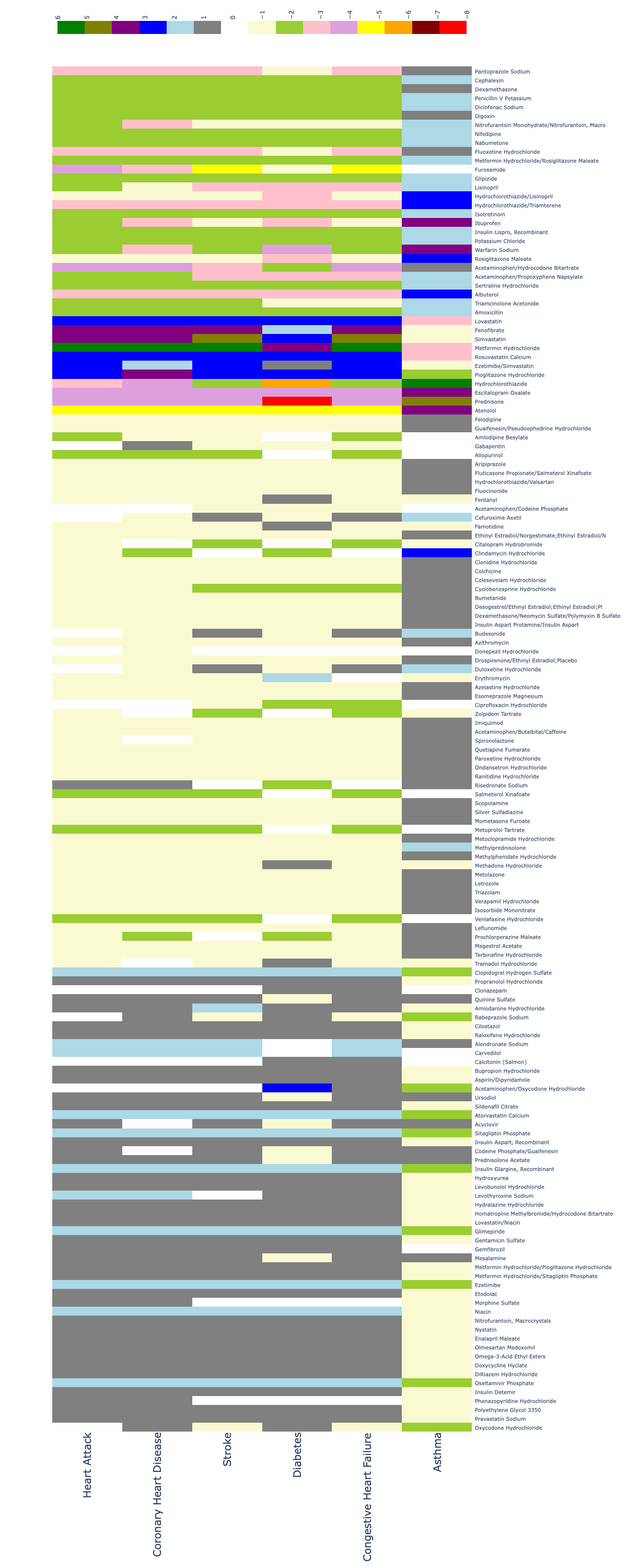}}
\caption{The heat map after bi-clustering, X-axis stands for the 6 disease conditions and Y-axis is the name of the 153 drugs that have an influence on any of the diseases involved in our experiment. The color bar above the heat map annotates the scores that different colors in the heat map stand for.}
\label{fig:BIC}
\end{figure}

\subsection*{Evaluation of Known Drug-disease Associations}

The results of the prediction at K for 6 disease conditions are shown in Figure \ref{fig:top-k}. The figure demonstrates the precision of our prediction at $K\in\{5,10,15,20\}$. For type 2 diabetes, stroke, heart attack and congestive heart failure, it is clear that most of the drugs can be mapped into the ground truth(SIDER drug list) when the K is small, the precisions of all the four disease conditions in the figure are greater than or equal to 0.8 when $K=5$, their precision will decrease with the increase of K. However, the results of asthma and coronary heart disease were not as expected. For coronary heart disease, there is not so many known drug-disease pair in SIDER, which could be a reason for the low precision of this disease condition. Some of the clinical variables, like cholesterol, LDL (low-density lipoprotein), HDL (high-density lipoprotein) and triglycerides are more salient features than other clinical variables. So, our analysis for the disease condition which does not have a strong correlation with these clinical variables could have low precision. Apart from those known treatments of each target disease that can be found in the ground truth, there could be some unknown drug candidates which are likely to be the treatment of target disease. 


The results of the FE test are shown in Figure \ref{fig:FEScore}. As we can see, there is a negative linear relationship between the FE score and the group order. Since the average FE test score is decreasing with the ascending order of groups, so there is a positive linear relationship between FE score and average repurposing possibility score. The result in Figure \ref{fig:FEScore} shows that all 6 disease conditions demonstrate a negative linear relationship between their FE score and group order. Therefore, our scoring function is proven to be reasonable.


\subsection*{Case Study and Explainability}
Having presented that our model successfully identified known associations between drugs and diseases, we further demonstrate the explainability of our model via corresponding complementarity between laboratory tests of drug effect vectors and disease sign vectors. To exemplify this, we select 5 drug-disease association pairs (i.e., Type 2 diabetes-Clopidogrel hydrogen sulfate, Type 2 diabetes-Doxycycline hyclate, Coronary heart disease-Alendronate sodium, Congestive heart failure-Alendronate sodium and Heart attack-Alendronate sodium). For a given disease, the selected drugs are in its top-20 predicted list but not have been indicated as the treatment. In order to vividly compare the clinical vectors of the drug candidates and the corresponding disease, we present the clinical variables which contribute to their repurposing possibility score in Table \ref{Tab:Cases}. All the detailed clinical vectors can be found in \textbf{Table S2} and \textbf{Table S3} of Supplementary Materials. Combining the clinical disease vectors with clinical drug effect vectors, we can analyze why the drug candidates we select are potential treatments for corresponding disease conditions from the standpoint of clinical variables included in our experiment. 


\begin{table*}[!h]
\caption{
This table presents the selected previously unknown drug-disease pairs predicted by our method, we just show the clinical variables that contribute to the final repurposing possibility of each drug-disease pair in the table.
}
\begin{tabular}{cccccccc}
\hline
Laboratory Test &       & Alkaline Phosphatase & Cholesterol& Glucose & HDL & LDL & Triglycerides \\ \hline
Disease &Type 2 Diabetes              & $\downarrow$      & $\uparrow$   & $\uparrow$    & $\downarrow$                   & $\uparrow$                    & $\uparrow$    \\ 
Drug &Clopidogrel Hydrogen Sulfate & - & $\downarrow$ & - & -                             & $\downarrow$                  & -             \\ \hline
Disease &Type 2 Diabetes              & $\downarrow$      & $\uparrow$   & $\uparrow$    & $\downarrow$                   & $\uparrow$                    & $\uparrow$    \\ 
Drug &Doxycycline Hyclate  & - &  -  & $\downarrow$  & -& - & - \\ \hline
Disease &Coronary Heart Disease       & $\uparrow$       & $\uparrow$   &$\uparrow$& $\downarrow$                   & $\uparrow$                    & $\uparrow$    \\ 
Drug &Alendronate Sodium           & $\downarrow$     & - &   -      & $\uparrow$                     & -                             & -             \\ \hline
Disease &Congestive Heart Failure     & $\uparrow$      & $\uparrow$& $\uparrow$  & $\downarrow$                   & $\uparrow$                    & $\uparrow$    \\ 
Drug &Alendronate Sodium           & $\downarrow$     & - &   -      & $\uparrow$                     & -                             & -             \\ \hline
Disease &Heart Attack                 & $\uparrow$      & $\uparrow$& $\uparrow$  & $\downarrow$                   & $\uparrow$                    & $\uparrow$  \\
Drug &Alendronate Sodium           & $\downarrow$     & - &   -      & $\uparrow$                     & -                             & -             \\ \hline
\end{tabular}
\label{Tab:Cases}
\end{table*}

In the case of type 2 diabetes, we found that clopidogrel hydrogen sulfate could have a therapeutic effect on type 2 diabetes and Doxycycline Hyclate. Clopidogrel hydrogen sulfate is an antiplatelet medication and can be used to reduce the risk of myocardial infarction and stroke \cite{jarvis2000clopidogrel}. A study reported that clopidogrel will alleviate insulin resistance and improve glycemic control in type 2 diabetic patients \cite{taher2011beneficial}, which is an important cause of insulin resistance. From the clinical drug effect vector of clopidogrel and clinical disease sign vector of type 2 diabetes, we can see clopidogrel and type 2 diabetes have the opposite effect on the cholesterol and LDL level. Lower cholesterol and LDL levels are biological markers of good glycaemic control \cite{khan2007association}, which is also corresponding to the literature study. 
Doxycycline Hyclate is an antibiotic which is primarily used to treat a wide range of bacterial infections. From the clinical vectors of Doxycycline and type 2 diabetes, we can see they have the opposite effect on the serum glucose level. High fasting blood glucose level is a common biological marker among type 2 diabetes patients. This finding is supported by a medical study that doxycycline can improve insulin resistance and fasting blood glucose level \cite{wang2017low}. 
The analysis based on the opposite effect of type 2 diabetes and clopidogrel proves our prediction is reasonable, clopidogrel and doxycycline may be used as treatments for type 2 diabetes . 


Alendronate sodium is usually used to treat osteoporosis \cite{porras1999pharmacokinetics}. We found it can potentially have a therapeutic effect on cardiovascular disease, including congestive heart failure, heart attack and coronary heart disease. Experiments show that alendronate can induce significantly lower cardiovascular mortality and reduce the risk of cardiovascular incidents \cite{sing2018association}. A possible explanation given by this study is that bone and cardiovascular remodeling share some biological markers. From the clinical drug effect vectors of alendronate, we can see alendronate can lower alkaline phosphatase (ALP) and elevate the HDL level. Researches show that ALP can catalyze the inhibitor of vascular calcification, thus high-level ALP may lead to vascular hardening and promotes the atherosclerotic process \cite{panh2017association}. On the other hand, HDL will promote reverse cholesterol transport, which could reduce the risk of cardiovascular events \cite{rader2014hdl}. Thus, it seems possible that alendronate could be repurposed as a treatment for cardiovascular disease. 

\subsection*{Highly related drugs and diseases}
In Fig. \ref{fig:DDS}, we demonstrate part of the repurposing possibility scores in the form of heat map. To further digging the relation within different drugs or diseases, we use bi-clustering algorithm to do a clustering for the drugs and diseases in Fig. \ref{fig:DDS}. Bi-clustering is a data mining technology that simultaneous clustering of both row and column sets in a data matrix \cite{mirkin2013mathematical}. Given an $m \times n$, bi-clustering algorithm will generate new $m \times n$ matrix that a subset of rows which exhibit similar behavior across a subset of columns, or vice versa. In our work, we use bi-clustering algorithm to find different drugs with similar effect on some a disease and different diseases which can be treated with same kind of drug. The clustering result is plotted in Fig. \ref{fig:BIC}. As shown in Fig. \ref{fig:BIC}, type 2 diabetes has a strong correlation with heart diseases and stroke. We can also find many drugs that can decrease blood lipid or sugar level have a therapy effect on those diseases. In the further, these findings can help to find potential drug-disease pairs.

\subsection*{Limitations and further work}
The verification results above show that our framework may identify some potential drug indications and thus help researchers find novel uses of existing drugs. However, our framework still has some limitations and space to improve. 

First of all, we only include 6 kinds of diseases and 392 kinds of drugs in the our work. Actually, there are some other disease conditions and drugs that can be found in NHANES and EHR dataset. The reason we just include a part of drugs and diseases is that many of them have a small sample size so that we can not get a reliable result from them. To guarantee the results we get from the dataset are reliable enough, the sample size of each drug and disease that included in this work is larger than 1000. Due to this threshold, the experiments are conducted on 6 diseases conditions and 392 drugs, but the results we get are reliable and robust. In the future, we can include more drug-disease pairs with a large-scaled dataset. 

The second limitation is the clinical variables involved in the experiment. Hundreds of clinical variables (laboratory tests) can be found in the NHANES dataset, but we still need to match them with the clinical variables in the EHR dataset. However, 35 kinds of clinical variables cannot completely reflect the human physiological activity, so it would be also addressed if we have a larger EHR dataset that contains more clinical variables. 


\section*{Conclusion}

In this paper, we establish a drug repurposing computational framework by using the electronic clinical information from the National Health and Nutrition Examination Survey (NHANES) and Electronic Health Records(EHR). We consider both of the opposite and same expressions between clinical disease sign vector and clinical drug effect vector in each drug-disease pair to calculate the repurposing possibility score. Our inferences of the novel use for different drugs are based on their repurposing possibility score with different disease conditions.  We verify our predictions by fold-enrichment test and top K precision. Then, we further prove the feasibility of our model by doing a literature analysis of our prediction result. The result shows that our framework can not only retrieve the known indications of existing drugs but also find the previously unknown indications of existing drugs. So our framework can be potentially used in the drug repurposing tasks.



\begin{backmatter}

\section*{Acknowledgements}
Not applicable.
  
\section*{Author's contributions}
PZ conceived the project. QW, RL, and PZ developed the method. QW conducted the experiments. QW, RL, and PZ analyzed experimental results. QW, RL, and PZ wrote the manuscript. All authors read and approved the final manuscript.
  
\section*{Funding}
This work was funded in part by the National Center for Advancing Translational Research of the National Institutes of Health under award number CTSA Grant UL1TR002733. The content is solely the responsibility of the authors and does not necessarily represent the official views of the National Institutes of Health.

\section*{Availability of data and materials}
\textcolor{black}{NHANES data analysed in the study is available on \href{https://wwwn.cdc.gov/nchs/nhanes/Default.aspx}{National Center for Health Statistics}. The source code is provided for reproducing and is available at \url{https://github.com/HoytWen/CCMDR}.}

\section*{Ethics approval and consent to participate}
Not applicable.

\section*{Consent for publication}
Not applicable.

\section*{Competing interests}
The authors declare that they have no competing interests.



\newcommand{\BMCxmlcomment}[1]{}

\BMCxmlcomment{

<refgrp>

<bibl id="B1">
  <title><p>Finding new tricks for old drugs: an efficient route for
  public-sector drug discovery</p></title>
  <aug>
    <au><snm>O'Connor</snm><fnm>KA</fnm></au>
    <au><snm>Roth</snm><fnm>BL</fnm></au>
  </aug>
  <source>Nature reviews Drug discovery</source>
  <publisher>Nature Publishing Group</publisher>
  <pubdate>2005</pubdate>
  <volume>4</volume>
  <issue>12</issue>
  <fpage>1005</fpage>
</bibl>

<bibl id="B2">
  <title><p>New uses for old drugs</p></title>
  <aug>
    <au><snm>Chong</snm><fnm>CR</fnm></au>
    <au><snm>Sullivan Jr</snm><fnm>DJ</fnm></au>
  </aug>
  <source>Nature</source>
  <publisher>Nature Publishing Group</publisher>
  <pubdate>2007</pubdate>
  <volume>448</volume>
  <issue>7154</issue>
  <fpage>645</fpage>
</bibl>

<bibl id="B3">
  <title><p>New drug development in the United States from 1963 to
  1999</p></title>
  <aug>
    <au><snm>DiMasi</snm><fnm>JA</fnm></au>
  </aug>
  <source>Clinical Pharmacology and Therapeutics</source>
  <publisher>Wiley Online Library</publisher>
  <pubdate>2001</pubdate>
  <volume>69</volume>
  <issue>5</issue>
  <fpage>286</fpage>
  <lpage>-296</lpage>
</bibl>

<bibl id="B4">
  <title><p>Estimating the cost of new drug development: is it really 802
  million?</p></title>
  <aug>
    <au><snm>Adams</snm><fnm>CP</fnm></au>
    <au><snm>Brantner</snm><fnm>VV</fnm></au>
  </aug>
  <source>Health affairs</source>
  <publisher>Project HOPE-The People-to-People Health Foundation,
  Inc.</publisher>
  <pubdate>2006</pubdate>
  <volume>25</volume>
  <issue>2</issue>
  <fpage>420</fpage>
  <lpage>-428</lpage>
</bibl>

<bibl id="B5">
  <title><p>Drug repurposing: progress, challenges and
  recommendations</p></title>
  <aug>
    <au><snm>Pushpakom</snm><fnm>S</fnm></au>
    <au><snm>Iorio</snm><fnm>F</fnm></au>
    <au><snm>Eyers</snm><fnm>PA</fnm></au>
    <au><snm>Escott</snm><fnm>KJ</fnm></au>
    <au><snm>Hopper</snm><fnm>S</fnm></au>
    <au><snm>Wells</snm><fnm>A</fnm></au>
    <au><snm>Doig</snm><fnm>A</fnm></au>
    <au><snm>Guilliams</snm><fnm>T</fnm></au>
    <au><snm>Latimer</snm><fnm>J</fnm></au>
    <au><snm>McNamee</snm><fnm>C</fnm></au>
    <au><cnm>others</cnm></au>
  </aug>
  <source>Nature reviews Drug discovery</source>
  <publisher>Nature Publishing Group</publisher>
  <pubdate>2019</pubdate>
  <volume>18</volume>
  <issue>1</issue>
  <fpage>41</fpage>
  <lpage>-58</lpage>
</bibl>

<bibl id="B6">
  <title><p>Predicting new molecular targets for known drugs</p></title>
  <aug>
    <au><snm>Keiser</snm><fnm>MJ</fnm></au>
    <au><snm>Setola</snm><fnm>V</fnm></au>
    <au><snm>Irwin</snm><fnm>JJ</fnm></au>
    <au><snm>Laggner</snm><fnm>C</fnm></au>
    <au><snm>Abbas</snm><fnm>AI</fnm></au>
    <au><snm>Hufeisen</snm><fnm>SJ</fnm></au>
    <au><snm>Jensen</snm><fnm>NH</fnm></au>
    <au><snm>Kuijer</snm><fnm>MB</fnm></au>
    <au><snm>Matos</snm><fnm>RC</fnm></au>
    <au><snm>Tran</snm><fnm>TB</fnm></au>
    <au><cnm>others</cnm></au>
  </aug>
  <source>Nature</source>
  <publisher>Nature Publishing Group</publisher>
  <pubdate>2009</pubdate>
  <volume>462</volume>
  <issue>7270</issue>
  <fpage>175</fpage>
</bibl>

<bibl id="B7">
  <title><p>The Connectivity Map: using gene-expression signatures to connect
  small molecules, genes, and disease</p></title>
  <aug>
    <au><snm>Lamb</snm><fnm>J</fnm></au>
    <au><snm>Crawford</snm><fnm>ED</fnm></au>
    <au><snm>Peck</snm><fnm>D</fnm></au>
    <au><snm>Modell</snm><fnm>JW</fnm></au>
    <au><snm>Blat</snm><fnm>IC</fnm></au>
    <au><snm>Wrobel</snm><fnm>MJ</fnm></au>
    <au><snm>Lerner</snm><fnm>J</fnm></au>
    <au><snm>Brunet</snm><fnm>JP</fnm></au>
    <au><snm>Subramanian</snm><fnm>A</fnm></au>
    <au><snm>Ross</snm><fnm>KN</fnm></au>
    <au><cnm>others</cnm></au>
  </aug>
  <source>science</source>
  <publisher>American Association for the Advancement of Science</publisher>
  <pubdate>2006</pubdate>
  <volume>313</volume>
  <issue>5795</issue>
  <fpage>1929</fpage>
  <lpage>-1935</lpage>
</bibl>

<bibl id="B8">
  <title><p>The Connectivity Map: a new tool for biomedical
  research</p></title>
  <aug>
    <au><snm>Lamb</snm><fnm>J</fnm></au>
  </aug>
  <source>Nature reviews cancer</source>
  <publisher>Nature Publishing Group</publisher>
  <pubdate>2007</pubdate>
  <volume>7</volume>
  <issue>1</issue>
  <fpage>54</fpage>
  <lpage>-60</lpage>
</bibl>

<bibl id="B9">
  <title><p>DPDR-CPI, a server that predicts drug positioning and drug
  repositioning via chemical-protein interactome</p></title>
  <aug>
    <au><snm>Luo</snm><fnm>H</fnm></au>
    <au><snm>Zhang</snm><fnm>P</fnm></au>
    <au><snm>Cao</snm><fnm>XH</fnm></au>
    <au><snm>Du</snm><fnm>D</fnm></au>
    <au><snm>Ye</snm><fnm>H</fnm></au>
    <au><snm>Huang</snm><fnm>H</fnm></au>
    <au><snm>Li</snm><fnm>C</fnm></au>
    <au><snm>Qin</snm><fnm>S</fnm></au>
    <au><snm>Wan</snm><fnm>C</fnm></au>
    <au><snm>Shi</snm><fnm>L</fnm></au>
    <au><cnm>others</cnm></au>
  </aug>
  <source>Scientific reports</source>
  <publisher>Nature Publishing Group</publisher>
  <pubdate>2016</pubdate>
  <volume>6</volume>
  <issue>1</issue>
  <fpage>1</fpage>
  <lpage>-9</lpage>
</bibl>

<bibl id="B10">
  <title><p>Drug repositioning by kernel-based integration of molecular
  structure, molecular activity, and phenotype data</p></title>
  <aug>
    <au><snm>Wang</snm><fnm>Y</fnm></au>
    <au><snm>Chen</snm><fnm>S</fnm></au>
    <au><snm>Deng</snm><fnm>N</fnm></au>
    <au><snm>Wang</snm><fnm>Y</fnm></au>
  </aug>
  <source>PloS one</source>
  <publisher>Public Library of Science</publisher>
  <pubdate>2013</pubdate>
  <volume>8</volume>
  <issue>11</issue>
</bibl>

<bibl id="B11">
  <title><p>Predicting drug-disease associations by using similarity
  constrained matrix factorization</p></title>
  <aug>
    <au><snm>Zhang</snm><fnm>W</fnm></au>
    <au><snm>Yue</snm><fnm>X</fnm></au>
    <au><snm>Lin</snm><fnm>W</fnm></au>
    <au><snm>Wu</snm><fnm>W</fnm></au>
    <au><snm>Liu</snm><fnm>R</fnm></au>
    <au><snm>Huang</snm><fnm>F</fnm></au>
    <au><snm>Liu</snm><fnm>F</fnm></au>
  </aug>
  <source>BMC bioinformatics</source>
  <publisher>BioMed Central</publisher>
  <pubdate>2018</pubdate>
  <volume>19</volume>
  <issue>1</issue>
  <fpage>1</fpage>
  <lpage>-12</lpage>
</bibl>

<bibl id="B12">
  <title><p>The productivity crisis in pharmaceutical RD</p></title>
  <aug>
    <au><snm>Pammolli</snm><fnm>F</fnm></au>
    <au><snm>Magazzini</snm><fnm>L</fnm></au>
    <au><snm>Riccaboni</snm><fnm>M</fnm></au>
  </aug>
  <source>Nature reviews Drug discovery</source>
  <publisher>Nature Publishing Group</publisher>
  <pubdate>2011</pubdate>
  <volume>10</volume>
  <issue>6</issue>
  <fpage>428</fpage>
  <lpage>-438</lpage>
</bibl>

<bibl id="B13">
  <title><p>Inferring disease association using clinical factors in a
  combinatorial manner and their use in drug repositioning</p></title>
  <aug>
    <au><snm>Jung</snm><fnm>J</fnm></au>
    <au><snm>Lee</snm><fnm>D</fnm></au>
  </aug>
  <source>Bioinformatics</source>
  <publisher>Oxford University Press</publisher>
  <pubdate>2013</pubdate>
  <volume>29</volume>
  <issue>16</issue>
  <fpage>2017</fpage>
  <lpage>-2023</lpage>
</bibl>

<bibl id="B14">
  <title><p>Inferring new drug indications using the complementarity between
  clinical disease signatures and drug effects</p></title>
  <aug>
    <au><snm>Jang</snm><fnm>D</fnm></au>
    <au><snm>Lee</snm><fnm>S</fnm></au>
    <au><snm>Lee</snm><fnm>J</fnm></au>
    <au><snm>Kim</snm><fnm>K</fnm></au>
    <au><snm>Lee</snm><fnm>D</fnm></au>
  </aug>
  <source>Journal of biomedical informatics</source>
  <publisher>Elsevier</publisher>
  <pubdate>2016</pubdate>
  <volume>59</volume>
  <fpage>248</fpage>
  <lpage>-257</lpage>
</bibl>

<bibl id="B15">
  <title><p>Computational drug repositioning using continuous self-controlled
  case series</p></title>
  <aug>
    <au><snm>Kuang</snm><fnm>Z</fnm></au>
    <au><snm>Thomson</snm><fnm>J</fnm></au>
    <au><snm>Caldwell</snm><fnm>M</fnm></au>
    <au><snm>Peissig</snm><fnm>P</fnm></au>
    <au><snm>Stewart</snm><fnm>R</fnm></au>
    <au><snm>Page</snm><fnm>D</fnm></au>
  </aug>
  <source>Proceedings of the 22nd ACM SIGKDD international conference on
  knowledge discovery and data mining</source>
  <pubdate>2016</pubdate>
  <fpage>491</fpage>
  <lpage>-500</lpage>
</bibl>

<bibl id="B16">
  <title><p>Exploiting electronic health records to mine drug effects on
  laboratory test results</p></title>
  <aug>
    <au><snm>Ghalwash</snm><fnm>M</fnm></au>
    <au><snm>Li</snm><fnm>Y</fnm></au>
    <au><snm>Zhang</snm><fnm>P</fnm></au>
    <au><snm>Hu</snm><fnm>J</fnm></au>
  </aug>
  <source>Proceedings of the 2017 ACM on Conference on Information and
  Knowledge Management</source>
  <pubdate>2017</pubdate>
  <fpage>1837</fpage>
  <lpage>-1846</lpage>
</bibl>

<bibl id="B17">
  <title><p>National Health and Nutrition Examination Survey. NCFHS
  (NCHS)</p></title>
  <aug>
    <au><snm>Cdc</snm><fnm>C</fnm></au>
  </aug>
  <source>US Department of Health and Human Services, Centers for Disease
  Control and Prevention</source>
  <pubdate>2005</pubdate>
</bibl>

<bibl id="B18">
  <title><p>Statistical significance for genomewide studies</p></title>
  <aug>
    <au><snm>Storey</snm><fnm>JD</fnm></au>
    <au><snm>Tibshirani</snm><fnm>R</fnm></au>
  </aug>
  <source>Proceedings of the National Academy of Sciences</source>
  <publisher>National Acad Sciences</publisher>
  <pubdate>2003</pubdate>
  <volume>100</volume>
  <issue>16</issue>
  <fpage>9440</fpage>
  <lpage>-9445</lpage>
</bibl>

<bibl id="B19">
  <title><p>Regression shrinkage and selection via the lasso</p></title>
  <aug>
    <au><snm>Tibshirani</snm><fnm>R</fnm></au>
  </aug>
  <source>Journal of the Royal Statistical Society: Series B
  (Methodological)</source>
  <publisher>Wiley Online Library</publisher>
  <pubdate>1996</pubdate>
  <volume>58</volume>
  <issue>1</issue>
  <fpage>267</fpage>
  <lpage>-288</lpage>
</bibl>

<bibl id="B20">
  <title><p>The SIDER database of drugs and side effects</p></title>
  <aug>
    <au><snm>Kuhn</snm><fnm>M</fnm></au>
    <au><snm>Letunic</snm><fnm>I</fnm></au>
    <au><snm>Jensen</snm><fnm>LJ</fnm></au>
    <au><snm>Bork</snm><fnm>P</fnm></au>
  </aug>
  <source>Nucleic acids research</source>
  <publisher>Oxford University Press</publisher>
  <pubdate>2015</pubdate>
  <volume>44</volume>
  <issue>D1</issue>
  <fpage>D1075</fpage>
  <lpage>-D1079</lpage>
</bibl>

<bibl id="B21">
  <title><p>Clopidogrel</p></title>
  <aug>
    <au><snm>Jarvis</snm><fnm>B</fnm></au>
    <au><snm>Simpson</snm><fnm>K</fnm></au>
  </aug>
  <source>Drugs</source>
  <publisher>Springer</publisher>
  <pubdate>2000</pubdate>
  <volume>60</volume>
  <issue>2</issue>
  <fpage>347</fpage>
  <lpage>-377</lpage>
</bibl>

<bibl id="B22">
  <title><p>Beneficial effects of clopidogrel on glycemic indices and oxidative
  stress in patients with type 2 diabetes</p></title>
  <aug>
    <au><snm>Taher</snm><fnm>MA</fnm></au>
    <au><snm>Nassir</snm><fnm>ES</fnm></au>
  </aug>
  <source>Saudi Pharmaceutical Journal</source>
  <publisher>Elsevier</publisher>
  <pubdate>2011</pubdate>
  <volume>19</volume>
  <issue>2</issue>
  <fpage>107</fpage>
  <lpage>-113</lpage>
</bibl>

<bibl id="B23">
  <title><p>Association between glycaemic control and serum lipids profile in
  type 2 diabetic patients: HbA 1c predicts dyslipidaemia</p></title>
  <aug>
    <au><snm>Khan</snm><fnm>HA</fnm></au>
    <au><snm>Sobki</snm><fnm>SH</fnm></au>
    <au><snm>Khan</snm><fnm>SA</fnm></au>
  </aug>
  <source>Clinical and experimental medicine</source>
  <publisher>Springer</publisher>
  <pubdate>2007</pubdate>
  <volume>7</volume>
  <issue>1</issue>
  <fpage>24</fpage>
  <lpage>-29</lpage>
</bibl>

<bibl id="B24">
  <title><p>Low dose doxycycline decreases systemic inflammation and improves
  glycemic control, lipid profiles, and islet morphology and function in db/db
  mice</p></title>
  <aug>
    <au><snm>Wang</snm><fnm>N</fnm></au>
    <au><snm>Tian</snm><fnm>X</fnm></au>
    <au><snm>Chen</snm><fnm>Y</fnm></au>
    <au><snm>Tan</snm><fnm>Hq</fnm></au>
    <au><snm>Xie</snm><fnm>Pj</fnm></au>
    <au><snm>Chen</snm><fnm>Sj</fnm></au>
    <au><snm>Fu</snm><fnm>Yc</fnm></au>
    <au><snm>Chen</snm><fnm>Yx</fnm></au>
    <au><snm>Xu</snm><fnm>Wc</fnm></au>
    <au><snm>Wei</snm><fnm>Cj</fnm></au>
  </aug>
  <source>Scientific reports</source>
  <publisher>Nature Publishing Group</publisher>
  <pubdate>2017</pubdate>
  <volume>7</volume>
  <issue>1</issue>
  <fpage>1</fpage>
  <lpage>-15</lpage>
</bibl>

<bibl id="B25">
  <title><p>Pharmacokinetics of alendronate</p></title>
  <aug>
    <au><snm>Porras</snm><fnm>AG</fnm></au>
    <au><snm>Holland</snm><fnm>SD</fnm></au>
    <au><snm>Gertz</snm><fnm>BJ</fnm></au>
  </aug>
  <source>Clinical pharmacokinetics</source>
  <publisher>Springer</publisher>
  <pubdate>1999</pubdate>
  <volume>36</volume>
  <issue>5</issue>
  <fpage>315</fpage>
  <lpage>-328</lpage>
</bibl>

<bibl id="B26">
  <title><p>Association of alendronate and risk of cardiovascular events in
  patients with hip fracture</p></title>
  <aug>
    <au><snm>Sing</snm><fnm>CW</fnm></au>
    <au><snm>Wong</snm><fnm>AY</fnm></au>
    <au><snm>Kiel</snm><fnm>DP</fnm></au>
    <au><snm>Cheung</snm><fnm>EY</fnm></au>
    <au><snm>Lam</snm><fnm>JK</fnm></au>
    <au><snm>Cheung</snm><fnm>TT</fnm></au>
    <au><snm>Chan</snm><fnm>EW</fnm></au>
    <au><snm>Kung</snm><fnm>AW</fnm></au>
    <au><snm>Wong</snm><fnm>IC</fnm></au>
    <au><snm>Cheung</snm><fnm>CL</fnm></au>
  </aug>
  <source>Journal of Bone and Mineral Research</source>
  <publisher>Wiley Online Library</publisher>
  <pubdate>2018</pubdate>
  <volume>33</volume>
  <issue>8</issue>
  <fpage>1422</fpage>
  <lpage>-1434</lpage>
</bibl>

<bibl id="B27">
  <title><p>Association between serum alkaline phosphatase and coronary artery
  calcification in a sample of primary cardiovascular prevention
  patients</p></title>
  <aug>
    <au><snm>Panh</snm><fnm>L</fnm></au>
    <au><snm>Ruidavets</snm><fnm>JB</fnm></au>
    <au><snm>Rousseau</snm><fnm>H</fnm></au>
    <au><snm>Petermann</snm><fnm>A</fnm></au>
    <au><snm>Bongard</snm><fnm>V</fnm></au>
    <au><snm>B{\'e}rard</snm><fnm>E</fnm></au>
    <au><snm>Taraszkiewicz</snm><fnm>D</fnm></au>
    <au><snm>Lairez</snm><fnm>O</fnm></au>
    <au><snm>Galinier</snm><fnm>M</fnm></au>
    <au><snm>Carri{\'e}</snm><fnm>D</fnm></au>
    <au><cnm>others</cnm></au>
  </aug>
  <source>Atherosclerosis</source>
  <publisher>Elsevier</publisher>
  <pubdate>2017</pubdate>
  <volume>260</volume>
  <fpage>81</fpage>
  <lpage>-86</lpage>
</bibl>

<bibl id="B28">
  <title><p>HDL and cardiovascular disease</p></title>
  <aug>
    <au><snm>Rader</snm><fnm>DJ</fnm></au>
    <au><snm>Hovingh</snm><fnm>GK</fnm></au>
  </aug>
  <source>The Lancet</source>
  <publisher>Elsevier</publisher>
  <pubdate>2014</pubdate>
  <volume>384</volume>
  <issue>9943</issue>
  <fpage>618</fpage>
  <lpage>-625</lpage>
</bibl>

<bibl id="B29">
  <title><p>Mathematical classification and clustering</p></title>
  <aug>
    <au><snm>Mirkin</snm><fnm>B</fnm></au>
  </aug>
  <publisher>Springer Science and Business Media</publisher>
  <pubdate>2013</pubdate>
  <volume>11</volume>
</bibl>

</refgrp>
} 








\section*{Additional Files}
  \subsection*{Table S1 --- Laboratory Test List}
    This table includes the name of 35 kinds of laboratory test involved in our experiments and their corresponding NHANES code. 
    (Supplementary Data.pdf)

  \subsection*{Table S2 --- Clinical Disease Sign Vector}
    This table presents 6 kinds of disease conditions involved in our experiment and their influences on the 35 kinds of laboratory tests.
    (DiseaseVector.csv)
    
    \subsection*{Table S3 --- Clinical Drug Effect Vector}
    This table  presents 392 kinds of existing drugs or drug combinations involved in our experiment and their influences on the 35 kinds of laboratory tests.
    (DrugVector.csv)
    
    \subsection*{Table S4 --- Repurposing Possibility Score}
    This table include the repurposing possibility score of each drug-disease pair in our experiment. 
    (Repurposing Possibility Score.csv)
    
    \subsection*{Figure S1 --- Disease Clinical Variable Statistics}
    The figure present number of diseases will increase (Up) or decrease (Down) the level of each clinical variables. X-axis is the name of each clinical variable, Y-axis is the number diseases. Blue bar stands for the "Up" relation, red bar stands for the "Down" relation.
    (Supplementary Data.pdf)
    
    \subsection*{Figure S2 --- Drug Clinical Variable Statistics}
    The number of drugs will increase (Up) or decrease (Down) the level of each clinical variables. X-axis is the name of each clinical variable, Y-axis is the number diseases. Blue bar stands for the "Up" relation, red bar stands for the "Down" relation.
    (Supplementary Data.pdf)
    
    \subsection*{Figure S3 --- Detailed Drug-Disease Heat Map}
    We transform the repurposing possibility score table into heat map and present it in this figure. This version includes the repurposing possibility scores of all the drug-disease pair. 
    (HeatMap(full).png)

\end{backmatter}
\end{document}